\documentclass[%
reprint,longbibliography,preprintnumbers,
nofootinbib,
 amsmath,amssymb,
 aps,
prb,
]{revtex4-2}
\pdfoutput=1
\usepackage[utf8]{inputenc}
\usepackage{flushend}
\usepackage{dcolumn}
\usepackage{bm}
\usepackage{balance}
\usepackage{amssymb}
\usepackage{appendix}

\usepackage[normalem]{ulem}

\usepackage[colorlinks = true,
            linkcolor = blue,
            urlcolor  = blue,
            citecolor =green,
            anchorcolor = blue]{hyperref}
\usepackage{verbatim}
\usepackage{color,ulem}
\usepackage[english]{babel}

\usepackage[utf8]{inputenc}
\input Starburst.fd
\newcommand*\initfamily{\usefont{U}{Starburst}{xl}{n}}\initfamily

\newcommand{\beq}{\begin{eqnarray}}
\newcommand{\eeq}{\end{eqnarray}}
\usepackage{amsmath}
\usepackage{tikz}
\usetikzlibrary{decorations.pathmorphing}
\usetikzlibrary{shapes.misc}
\tikzset{cross/.style={cross out, draw=black, minimum size=8*(#1-\pgflinewidth), inner sep=0pt, outer sep=0pt},
cross/.default={1pt}}
\usetikzlibrary{patterns,math}
\begin{document}

\title{Microscopic mechanism of electric field-induced  superconductivity suppression in metallic thin films}

\author{Alessio Zaccone$^{1}$}
\author{Giovanni Alberto Ummarino$^{2}$}
\author{Alessandro Braggio$^{3}$}
\author{Francesco Giazotto$^{3}$}

\affiliation{$^1$  Department of Physics ``A. Pontremoli'', University of Milan, via Celoria 16,
20133 Milan, Italy}
\affiliation{$^2$ 
Politecnico di Torino, Dipartimento di Scienza Applicata e Tecnologia, corso Duca degli Abruzzi 24, 10129 Torino, Italy}
\affiliation{$^3$ 
NEST, Istituto Nanoscienze-CNR and Scuola Normale Superiore,  I-56127 Pisa, Italy}

 \vspace{1cm}

\begin{abstract}
Supercurrent field-effect transistors made from thin metallic films are a promising option for next-generation high-performance computation platforms. Despite extensive research, there is still no complete quantitative microscopic explanation for how an external DC electric field suppresses superconductivity in thin films. This study aims to provide a quantitative description of superconductivity as a function of film thickness based on Eliashberg's theory. The calculation considers the electrostatics of the electric field, its realistic penetration depth in the film, and its effect on the Cooper pair, which is described as a standard s-wave bound state according to BCS theory. The estimation suggests that an external electric field of approximately $10^8$
V/m is required to suppress superconductivity in 10-30-nm-thick films, which aligns with experimental observations. Ultimately, the study offers "materials by design" guidelines for suppressing supercurrent when an external electric field is applied to the film surface. Furthermore, the proposed framework can be easily extended to investigate the same effects for ultrathin films.
\end{abstract}

\maketitle

\section{Introduction}
Supercurrent field-effect transistors have massive potential for future classical \cite{Mukhanov04,Tolpygo16} and quantum computation \cite{Gambetta17} nanodevices. This is due to the recent experimentally discovery that superconductivity in thin metallic films can be suppressed by externally applying a strong enough external DC electric field, denoted as $E_{cr,ext}$ \cite{Giazotto1,Giazotto2,paolucci2019magnetotransport,golokolenov2021origin,paolucci2019field,elalaily2021gate,rocci2020gate,basset2021gate,alegria2021high,bours2020unveiling,rocci2020large,ritter2021superconducting,paolucci2019field2,catto2022microwave,de2019josephson,yu2023gate,de2020niobium,du2023high,puglia2020electrostatic,elalaily2023signatures,paolucci2023gate,ruf2023effects,ruf2024high,orus2021critical,koch2024gate}. Experimental evidence has shown that external electric fields (EF) on the order of $\sim 10^8$ V/m are required to suppress the supercurrent in metallic thin films with a thickness of around 20 nm \cite{review}. Despite several recent theoretical approaches in the literature \cite{Solinas,mercaldo2020electrically,virtanen2019superconducting,chirolli2021impact,mercaldo2021spectroscopic,amoretti2022destroying,Fomin,chakraborty2023microscopic}, the debate is still open and a quantitative microscopic prediction of the measured EF values $E_{cr,ext}$ needed to suppress the superconductivity in thin films experimentally is still missing.

This paper presents an internally consistent mechanism for the supercurrent field effect in metallic thin films by quantitatively predicting the EF magnitude needed to suppress superconductivity in alignment with experimental values. It is grounded in Eliashberg's microscopic theory and considers the electrostatic aspects of EF penetration and its screening.
For the first time, this theory can predict the critical field required to eliminate superconductivity in 10-30 nm-thick thin films without requiring any adjustable parameters. This prediction matches quite closely with experimentally measured values.
This new insight may provide material-design guidelines to optimize the supercurrent field effect. The impacts of structural disorder, sample geometry and thickness, and dielectric screening are quantitatively evaluated to fully predict the expected performance in gating of supercurrent-field effect transistors.

\section{Model assumptions}
\begin{figure}
\includegraphics[width = 1.0 \linewidth]{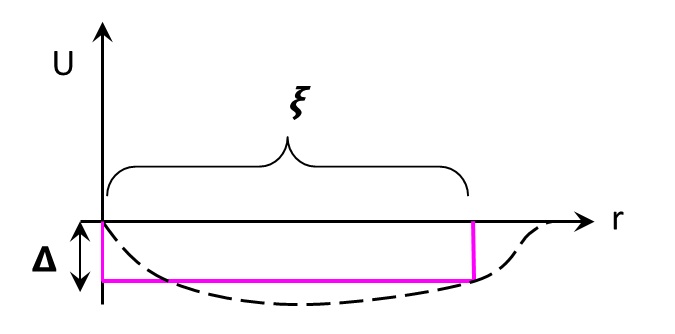}
	\caption{Schematic of the potential energy profile experienced by an electron bound in a Cooper pair (in the absence of external fields), where $r$ measures the radial distance from the other electron in the pair, following the original real-space description of Cooper pairs given by Weisskopf \cite{Weisskopf}. Therefore, the electron in a Cooper pair is effectively in a bound state of s-wave type. Applying a DC electric field to this s-wave bound state leads to field-induced electron tunneling, which escapes from the bound state with a probability $w$ outlined in textbooks \cite{Landau}. }
		\label{Weiss}
\end{figure}

The model is entirely generic for metallic thin films and relies on the following physical assumptions:\\
(i) The metallic thin film is thin enough that its penetration depth is comparable to the thickness or at least no less than an order of magnitude smaller. For example, with NbN thin films, one has a penetration depth of 4-5 nm \cite{Piatti}, comparable to the film thickness of $\sim 10-30$ nm. This ensures that the EF, exponentially decaying into the film, is never identically zero inside the film (recall that an exponentially decaying function is identically zero only for $r \rightarrow \infty$);\\
(ii) Even a minimal local value of the EF can tilt the attractive potential, which keeps a Cooper pair together \cite{Landau,Cooper}; \\
(iii) Following Cooper \cite{Cooper} and Weisskopf \cite{Weisskopf}, we assume the attractive potential that keeps the two electrons bound in the Cooper pair to be a spherically symmetric well (s-wave bound state), cf. Fig. \ref{Weiss};\\
(iv) Then the Cooper pair, according to BCS theory \cite{Cooper,BCS}, is governed by a Schr\"{o}dinger equation for an s-wave bound state. In the presence of a small but finite electric field pointing along the (arbitrary) direction $z$, the stationary Schr\"{o}dinger equation reads as \cite{Landau}:
\begin{equation}
    \left(\frac{1}{2}\nabla^{2} + \mathcal{E}+u(r) - E z\right) \psi = 0 \label{Landau}
\end{equation}
where $\mathcal{E}$ is the energy, $E$ is the electric field magnitude, $z$ is the spatial coordinate along which the EF points and $\psi$ is the wavefunction. In the above equation, atomic units are used. Furthermore, the attractive potential is schematically given by a spherical well: $u(r)=-\Delta$ for $0 \leq r \leq \xi$ and zero otherwise, where $\Delta$ is the BCS energy gap and $\xi$ is the coherence length (cf. Fig. \ref{Weiss}). The solutions to Eq. \eqref{Landau} are obtained by separating variables in parabolic coordinates and can be found in textbooks  \cite{Landau}. 
From the solution to Eq. \eqref{Landau}, one obtains the characteristic critical electric field magnitude $E_{cr}$ to break the Cooper pair;\\
(v) We also assume that the coherence length $\xi$ is smaller than the film thickness or, at most, comparable to the film thickness. This is because we treat the response of the superconducting phase as a whole; note that in experimentally realized thin films, the granularity and disorder could set the value of $\xi$ much smaller than in a bulk crystalline material. As we shall see, $\xi$ is also essential to compute the critical field $E_{cr}$.

The above assumptions can be applied to various materials and devices studied in the literature \cite{review}. For all these materials, the measured values of the critical field $E_{cr}$ always fall within the interval $\sim 1 \times 10^8$ V/m to $\sim 8 \times 10^8$ V/m.

\section{Structure of the paper}
In the following, we first summarize the microscopic Eliashberg theory framework used to quantitatively estimate the gap energy ($\Delta$) for actual materials, which corresponds to the binding energy of Cooper pairs. Next, we mathematically describe the two electrons in the Cooper pair as being in an s-wave bound state subjected to an electric field. We then revisit the derivation of the field-assisted tunneling probability, which indicates how the s-wave bound state dissociates (depairing probability). 
 From this solution, we derive an expression for the critical or characteristic value of the electric field (EF), $E_{cr}$, necessary to dissociate the Cooper pair as a function of $\Delta$ and the coherence length $\xi$ (which can be replaced by the mean free path $\ell$ in very disordered films). In the last part, we estimate the correction to the critical field due to the screening of the EF within the film thickness, examining the spatial decay of the EF from the surface into the film's interior. Accordingly, we then calculate the actual value of the externally applied EF that should be applied to induce the suppression of superconductivity in the film, $E_{cr,ext}$. The estimate is compared with experimental data for the illustrative case of NbN thin films.

\section{Quantitative theory of superconductivity in metallic thin films}

\subsection{Eliashberg equations}

We resort to the Eliashberg theory to compute a given material's BCS energy gap and the coherence length.
The standard (infinite) one-band s-wave Eliashberg equations, when the Migdal theorem holds \cite{UmmaMig}, are given in terms of the renormalization function $Z(i\omega_n)$ and the gap function $\Delta(i\omega_n)$ as \cite{revcarbi, revcarbimarsi, Allen, Parks, Marsiglio,ummarinorev,margine1}:
%
\begin{align}
\Delta(i\omega_n)Z(i\omega_n)&=\pi T\sum_{\omega_{n'}} \frac{\Delta(i\omega_{n'})}{\sqrt{\omega_{n'}^2+\Delta^{2}(i\omega_{n'})}} \\
&\times \big[ \lambda (i\omega_{n'}-i\omega_n)-\mu^{*}(\omega_{c})\theta(\omega_{c}-|\omega_{n'}|)\big]\nonumber\\
Z(i\omega_n)&=1+\frac{\pi T}{\omega_n}\sum_{\omega_{n'}}\frac{\omega_{n'}}{\sqrt{\omega_{n'}^2+\Delta^{2}(i\omega_{n'})}}\lambda (i\omega_{n'}-i\omega_n)
\end{align}
where $\theta(\omega_{c}-|\omega_{n'}|)$ is the Heaviside function, $\omega_{c}$ is a cut-off energy ($\omega_{c}> 3\Omega_{max}$, where $\Omega_{max}$ is the maximum phonon energy) \cite{Allen}, $\mu^{*}(\omega_{c})$ is the Coulomb pseudopotential,
and $\lambda (i\omega_{n'}-i\omega_n)$ is a function related to the electron-boson spectral density $\alpha^2F(\Omega)$ through the relation
\begin{equation} \label{lambda}
\lambda (i\omega_n'-i\omega_n)=2\int_0^\infty \frac{\Omega\,\alpha^2F(\Omega)}{\Omega^2+(\omega_{n'}-\omega_n)^2}d\Omega.
\end{equation}
The strength of the electron-phonon coupling intensity is given by the electron-phonon coupling parameter $\lambda=2\int_0^\infty \frac{\alpha^2F(\Omega) \,d\Omega}{\Omega}$.

The Eliashberg equations are usually solved numerically using an iterative method until numerical convergence is achieved.
The numerical procedure is simple when formulating it on the imaginary axis, but much less so for the real axis.
The critical temperature $T_{c}$ can be calculated either by solving an eigenvalue equation or, more efficiently, by assigning a minimal test value to the superconducting gap. For example, for Pb, it is $\Delta=1.34$ meV at $T=0$ K; then we fix $\Delta(T^*)=10^{-10}$ meV and compute the temperature $T^*$ at which the solution converges to find $T_c=T^*$. This method achieves a precision in the $T_{c}$ value significantly higher than the experimental confidence interval.

Suppose one removes the approximations of the infinite bandwidth, and takes the DOS equal to a constant (i.e., its value at the Fermi level). This situation arises when we consider the effect of the thickness in a thin film \cite{nostro} (see Appendix A). In that case, we can still solve the problem considering the quantum confinement correction for thin films, but the Eliashberg equations are slightly more complex, becoming four equations \cite{Allen}.
However, assuming the DOS to be symmetrical with respect to the Fermi level, $N(\varepsilon)=N(-\varepsilon)$, the situation simplifies again because the non-zero self-energy terms remain only two, i.e., just a generalization of the  equations discussed before for $Z(i\omega_n)$ and $\Delta(i\omega_n)Z(i\omega_n)$. In such case they read as \cite{carbin1,carbin2}
\begin{widetext}
\begin{align}
\Delta(i\omega_n)Z(i\omega_n)=&\pi T\sum_{\omega_{n'}} \frac{\Delta(i\omega_{n'})}{\sqrt{\omega_{n'}^2+\Delta^{2}(i\omega_{n'})}}[\frac{N(i\omega_{n'})+N(-i\omega_{n'})}{2}]\times\nonumber
\\
&\big[\lambda(i\omega_{n'}-i\omega_n)-\mu^{*}(\omega_c)\theta(\omega_c-|\omega_{n'}|)\big]\frac{2}{\pi} \arctan(\frac{W}{2Z(i\omega_{n'})\sqrt{\omega_{n'}^{2}+\Delta^{2}(i\omega_{n'})}})
\\
Z(i\omega_n)=& 1+\frac{\pi T}{\omega_n}\sum_{\omega_{n'}} \frac{\omega_{n'}}{\sqrt{\omega_{n'}^2+\Delta^{2}(i\omega_{n'})}}[\frac{N(i\omega_{n'})+N(-i\omega_{n'})}{2}]
\lambda (i\omega_{n'}-i\omega_n)\times\nonumber
\\
&\frac{2}{\pi}\arctan(\frac{W}{2Z(i\omega_{n'})\sqrt{\omega_{n'}^{2}+\Delta^{2}(i\omega_{n'})}})
\end{align}
\end{widetext}
where $N(\pm i\omega_{n'})=N(\pm Z(i\omega_{n'})\sqrt{(\omega_{n'})^{2}+\Delta^{2}(i\omega_{n'})})$ and the bandwidth $W$ is equal to half the Fermi energy, $E_{F}/2$.
Furthermore a symmetric DOS is also beneficial for getting faster and simpler numerical convergence. 


\subsection{Comparison with experimental data for the $T_c$}
The theory above has recently been shown to provide a quantitative, parameter-free prediction of the superconducting critical temperature in actual thin-film materials, such as Pb and Al thin films \cite{nostro}. 
We can now compare this theory's predicted values of $T_c$ with experimental data, particularly for NbN thin films. The Eliashberg equations mentioned earlier are solved using the experimentally determined Eliashberg spectrum $\alpha^{2}F(\Omega)$ found in Ref. \cite{Kihlstrom}, along with the concentration of free carriers as a function of film thickness, as reported in the literature \cite{Zhang}. This leaves no adjustable parameters for comparison.
The theoretical predictions are compared with experimental data from Ref. \cite{Zhang}. Without any adjustable parameters, the theory aligns excellently in quantitative terms with the experimental data for $L=20$ nm, predicting a $T_c$ value of 16.1 K. In what follows, we will concentrate our estimates on films of this thickness or similar. 

\section{Electric-field induced Cooper pair splitting}

The issue of splitting a Cooper pair by an electric field is similar to the classic problem of electric field-induced dissociation of an s-wave bound state. 
This similarity arises because, within BCS theory, a Cooper pair is represented as an s-wave bound state that satisfies the Schrödinger equation for two electrons interacting through an effective attractive force \cite{Cooper,BCS}, with a real space description initially proposed by Weisskopf and illustrated schematically in Fig. 
\ref{Weiss}. 
The solution for bound-state dissociation in an electric field is well known \cite{Landau} and has already been applied to the context of Cooper pair splitting by an external electric field in Refs. \cite{zaccone2023theory, Patino2021}.
The formula for the critical electric field magnitude needed to split the Cooper pair is given by:
\begin{equation}
E_{cr}=\frac{2 \Delta}{e\,\xi}, \label{crit}
\end{equation}
where $\Delta$ is the energy gap obtained by solution of Eliashberg equations, $e$ is the electron charge, and $\xi$ is the coherence length, which is self-consistently obtained by the solution of Eliashberg equations (Sec.IV and \cite{Carbotte}):
\begin{equation}
\xi(T)=\frac{v_F}{2}\frac{\sum_{n}\frac{\Delta^{2}(i\omega_n)}{Z(i\omega_n)[\omega^{2}_n+\Delta^{2}(i\omega_n)]^{1.5}}}{\sum_{n}\frac{\Delta^{2}(i\omega_n)}{\omega^{2}_n+\Delta^{2}(i\omega_n)}}.
\end{equation}

The above estimate for $E_{cr}$ can be obtained by considering Eq. \eqref{Landau}, i.e., the Schr\"{o}dinger equation for an electron initially bound in a s-wave bound state (the Cooper pair) of energy depth $\Delta$, and subjected to an external electric field of magnitude $E = |- \nabla V|$. 
While the s-wave bound state (the Cooper pair) is spherically symmetric, the electric field is directed along a specific spatial direction, which could be any direction in the solid angle. In analogy to what is shown in the textbooks, pp. 296-297 of Ref. \cite{Landau}, one solves the 
Schr\"{o}dinger equation Eq. \eqref{Landau} in parabolic coordinates, and uses the solution to compute the probability current of the electron escaping away from the bound state in the direction of the EF (i.e., the coordinate $z$ in Eq. \eqref{Landau}). 
The result for the probability $w$ of the electron tunneling away from a bound state, is given in dimensionless form as \cite{Landau}:
\begin{equation}
    w \sim \exp\left(-\frac{2}{3 E}\right),
\end{equation}
where $E$ is the non-dimensionalized electric field's magnitude (absolute value). For a s-wave bound state of unitary depth energy and unitary radius, converting to a dimensionful form, the above formula from \cite{Landau} reads as:
\begin{equation}
    w \sim \exp\left(-\frac{2}{3}\frac{E_a} {E}\right) \label{esc}
\end{equation}
where $E_{a}=2 R_H/e a_0$, with $R_H$ the Rydberg energy and $a_0$ the Bohr radius.
In particular, one finds that the critical electric field scale to dissociate the bound state is given by 
\begin{equation}
    E_{cr}=\frac{2 R_H}{e a_0}.
\end{equation}
The same result is valid for a generic bound state of depth energy $\Delta$ and range $\xi$, leding to \cite{Fomin}:
\begin{equation}
    E_{cr} = \frac{2 \Delta}{e \xi}.
\end{equation}
It is important to note that the direction $z$ of the electric field in Eq. \eqref{Landau} does not affect the critical value $E_{cr}$, since only the absolute value of the electric field, $E = |- \nabla V|$, appears in the solution for the escape probability\cite{Landau}

In the above equation, the spatial extent or range of the bound state is denoted with $\xi$. For the case of a Cooper pair, this is equal to the coherence length \cite{Weisskopf}, cf. Fig. \ref{Weiss}.
The coherence length ($\xi$) may depend on the material and is given by \cite{deGennes}:
\begin{equation}
    \frac{1}{\xi}=\frac{1}{\xi_0}+\frac{1}{\ell},\label{matt}
\end{equation}
where $\xi_0$ is the intrinsic (Pippard) coherence length, and $\ell$ is the mean free path. Thin films, such as those used in the supercurrent field effect devices, have a microstructure characterized by microcrystallites, that limit the values of $\ell$. Since, typically, $\ell \ll \xi_0$ (because $\xi_0$ can be tens or hundreds of nanometers), the coherence length $\xi$ is controlled by $\ell$, and, hence, by the disorder, and $\xi \approx \ell$. For experimental metallic thin film systems, the disorder is indeed always  present \cite{Sidorova}, and we can safely assume $\xi \approx \ell$. Therefore:
\begin{equation}
E_{cr}=\frac{2 \Delta}{e\,\ell}. \label{cr}
\end{equation}
Being in the diffusive regime, the coherence length $\ell$ has a small value, not that of a bulk superconductor but that of a thin film.
From our theory presented in Sect. IV.A-B, we know the gap energy $\Delta$ for a given material, and we can estimate the critical electric field $E_{cr}$ for superconductivity suppression inside the film under the assumptions stated in Sec. II.


For the example of NbN, the values of the critical electric field needed to suppress the superconductivity in 10-30 nm-thick thin films are of the order of $10^7$ V/m, under the assumption of no screening. More precisely, at $0 < T < 8$ K, the value, calculated based on Eq. \eqref{cr} with $\ell = 3.96$ $\textrm{\AA}$ and $\Delta$ evaluated from the Eliashberg theory, is practically constant with temperature, and equal to $E_{cr} = 1.4 \times 10^7$ V/m. 
This estimate is one order of magnitude lower than the experimental values of $\sim 10^8$ V/m reported in the literature for films of comparable thickness \cite{review}. 

As mentioned, this estimate still assumes perfect external electric field penetration to the sample. In other words, it does not consider the screening of the EF within the sample. The screening effects must be taken into account in order to determine the magnitude of the external EF required to suppress superconductivity, a process detailed in the next section, Sec. VI.

\section{Critical assessment of the electric field required to inhibit superconductivity in thin films}

\subsection{Average field within the sample}
In our theoretical estimate of the critical electric field $E_{cr}$ required to suppress superconductivity, as presented in the previous section, we implicitly assumed that the strength of the electric field (EF) acting to separate the Cooper pairs inside the film is identical to that of the externally applied electric field. However, it is clear that the magnitude of the EF within the thin film is significantly lower than that of the external EF and exhibits spatial heterogeneity due to the penetration profile. We will address these issues to develop a comprehensive quantitative prediction for superconductivity suppression throughout the thin film.

The EF magnitude in the thin film is significantly lower than the external EF due to screening. Here, we evaluate this reduction, which allows us to estimate $E_{cr,ext}$, the externally applied EF magnitude necessary to suppress superconductivity. 

Regarding the penetration of static and quasistatic electric fields into the superconducting phase of the material--and especially the temperature dependence thereof—-this remains an open issue in our understanding of superconductivity \cite{Waldram,Dressel,Vinokur,Hirsch,Tajmar,Salasnich,Diamantini2020,Mercereau,Testardi,balls}, which we cannot address 
here. 

In place of a conclusive theory, which would typically require, at least, the simultaneous numerical solution of several coupled differential equations, the Schr{\"o}dinger equation, the Ginzburg-Landau equation, the Maxwell equations, and the Poisson equation \cite{Lipavsky,Blatter,Virtanen,Chirolli}, we use here empirical input from ab-initio simulations and experiments of Piatti et al. \cite{Piatti}, for the same example of NbN thin films discussed above. 
According to that reference, the penetration depth of the electric field into NbN superconducting thin films is significantly greater than what is predicted by Thomas-Fermi estimates and other standard screening theories of (normal) metals, which typically measure just a few Angstroms in the range from 0.5 to 3 nm.

According to the screening theory of normal metals \cite{Jackson}, see also \cite{Ullman,Welser,Saslow}, the electric field incident onto the surface of a metal decays exponentially from the value it has at the interface. Let us denote with $E_{0}$ the value of EF at the interface between the metal and the surrounding environment and $x=0$ the interface coordinate. Hence, $E_{0}\equiv E(x=0)$. We assume, for simplicity, that $E_{0}$ coincides with the magnitude of the external EF as determined in the experiment. 
We use a common semi-empirical form for the profile of the EF inside the sample from Ref. \cite{Ullman}, which finds its justification from electrodynamics \cite{Jackson,Salasnich}:
\begin{equation}
E(x)=E_0 \exp(- x/l),
\end{equation}
where $l$ is a characteristic penetration length scale. For NbN in the superconducting state, this length is anomalously large (compared to normal metals), i.e., $l \approx 0.5-3$ nm, as shown in  \cite{Piatti}.
In agreement with the experimental protocol \cite{Giazotto1,Giazotto2}, we shall assume a bipolar setup whereby the EF is incident on both sides (surfaces) of the film. This implies that the EF is exponentially decaying inside the film from its external value $E_0$, on both sides at $z=0$ and $z=L$ \cite{Amoretti_2023}. Hence, we have complete symmetry across the midline $z=L/2$ plane of bilateral symmetry. This simplifies our problem as we only need to consider the exponential decay on one side of the film. We can safely assume that the superconductor "sees" the inhomogeneous electric field as mediated over the scale of, at least, $\xi$. However, the thickness of the film is assumed not to be more than the coherence length. We can thus estimate the "effective" electric field as the average magnitude of the EF inside the whole sample as
\begin{equation}
\bar{E}=\frac{1}{L/2}\int_{0}^{L/2}E_0\, e^{-x/l} dx = E_0 \frac{2l}{L}(1- e^{-L/(2l)}). \label{decay}
\end{equation}
In this formula, $E_0$ represents the external EF incident onto the film surface, while $\bar{E}$ represents the average value of the EF seen by the superconductor inside the film.

\subsection{Magnitude of the external electric field required to suppress superconductivity}
Using the empirical data from \cite{Piatti} for the example of NbN, we select a screening length of $\ell = 1 $ nm. With this choice, we can estimate the magnitude of the electric field EF that must be applied externally to break the Cooper pairs in the presence of dielectric screening within the film. The magnitude of EF must be sufficient to ensure that the average electric field $\bar{E}$ inside the sample equals the critical value $E_{cr}$, which is necessary to break the Cooper pairs. This critical value is calculated using Eq. \eqref{cr} and Eliashberg's theory discussed in Section V.
Hence, we must set $\bar{E}$ in Eq. \eqref{decay} equal to $E_{cr}$ evaluated in Sec. V, and solve for $E_0$, the value we supply externally to suppress the superconductivity. Let us call this value $E_0 \equiv E_{cr,ext}$ to distinguish it from $E_{cr}$ used before.
The magnitude of EF that has to be supplied externally is thus given by
\begin{equation}
E_{cr,ext}=\frac{E_{cr}} { \frac{2l}{L}(1- e^{-L/(2l)})}= \frac{2 \Delta/e\,\ell} {\frac{2l}{L}(1- e^{-L/(2l)})}. \label{extern}
\end{equation}
This equation represents one of the most important results of this paper. It establishes a direct quantitative relationship between the externally applied electric field magnitude, $E_{cr,ext}$, and several physical parameters that are specific to the material and the sample: $\Delta$ (material-dependent), $\ell$ (material-dependent), $l$ (material-dependent), and $L$ (sample geometry).

The final result for the EF magnitude, $E_{cr}$, that has to be supplied externally to cause the suppression of superconductivity in a $L=20$ nm thick NbN thin film, calculated based on Eq. \eqref{extern} with $l=1$ nm \cite{Piatti,Gonnelli}, is:
\begin{equation}
    E_{cr,ext} = 1.45 \times 10^8 \,\,\,\, \textrm{V/m}.
\end{equation}
We recall that, to obtain this value, the energy gap $\Delta$ in Eq. \eqref{extern} was computed in Sec. IV through Eliashberg's theory for NbN using an Eliashberg spectral function $\alpha^2F(\Omega)$ obtained experimentally. The latter includes structural disorder effects which are ubiquitous in experimental thin films.

It turns out that the value of the EF to be supplied is about $\sim 10^8$ V/m, which is in excellent agreement with the values measured experimentally \cite{review}. Choosing a lower value of screening length $l$ such as $l < 1$ nm within the range reported in \cite{Piatti,Gonnelli}, would lead to values of the critical electric field somewhat more significant than $1 \times 10^{8}$ V/m but still of the same order of magnitude and ideally in line with the experimental data in the literature \cite{review}.
In the above estimate, we have utilized a screening length value, $l$, that is temperature-independent. We will postpone discussing the trend of $E_{cr}$ versus $T$ until future studies when a comprehensive theory of the temperature dependence of $l$ in the superconducting phase be available.

\section{General mechanism of supercurrent suppression over the whole sample}
\subsection{Bipolarity}

The mechanism and calculations described above are "bipolar" in that they do not depend on the polarity of the applied electric field as long as a non-zero local EF is acting on the Cooper pair. According to Eq. \eqref{Landau}, the EF's local direction ($z$) is irrelevant, given that the s-wave bound state is spherically symmetric.
In our mathematical model, we assumed that the density of states is symmetrical concerning the Fermi level, which makes the calculations perfectly bipolar. This assumption approximates good metals where the Fermi level is relatively high due to the flattening of the Fermi square-root DOS at higher energies. Additionally, this approximation simplifies the numerical solutions to the Eliashberg equations, significantly reducing computational time.
If we had not made this assumption, a slightly asymmetric DOS would have resulted in a minor deviation from bipolarity in the mechanism. However, this effect would be negligible for metals, especially compared to experimental data. Nonetheless, this is an interesting point that we plan to revisit in future studies.

\subsection{Effects on the resistivity}
The investigation of the effects of an electrostatic field on superconductors is a longstanding issue \cite{glover}. In the normal state, conductivity can be weakly modified, and regarding the superconducting phase, the transition temperature can be either (weakly) positively or negatively shifted. 
Highly accurate measurements taken in the superconducting transition region for indium films \cite{glover} consistently showed that negative charging (i.e., adding electrons) leads to an increase in resistance, while positive charging (i.e., removing electrons) results in a corresponding decrease in resistance. When subjected to an electrostatic field of approximately $10^{7}$ V/m, the effects are minimal, producing a shift in transition temperature on the order of $10^{-4}$ K. 

This effect can be explained using a generalization of the proximity effect in Eliashberg's theory \cite{proxumma1,proxumma2}, employing a model with no free parameters \cite{indiumumma}. In all the cases examined (i.e., lead \cite{leadumma}, indium \cite{indiumumma}, and magnesium diboride \cite{MgB2umma}), the effect is minimal. In Eliashberg's theory, the relevant parameter is the electron-phonon spectral function $\alpha^{2}F(\Omega)$, which displays very weak dependence on the presence of electrostatic fields at this intensity. The same holds in Allen's theory \cite{allen1978}, which is used to calculate the resistivity as a function of temperature and is connected to Eliashberg's theory. In this context, the electron-phonon transport spectral function $\alpha^{2}F_{tr}(\Omega)$ emerges, which is closely related to the Eliashberg spectral function $\alpha^{2}F(\Omega)$ in the superconducting state \cite{allen1978}. Likewise, in this scenario, electrostatic fields of this intensity alter the electron-phonon transport spectral function almost imperceptibly; thus, the effects are also minimal. Since these two functions are also linked to the value of the DOS at the Fermi level, the minor variations in critical temperature and resistivity depend on the direction of the applied electrostatic field, as the normal DOS may be asymmetric relative to the Fermi level.

\section{"Materials by design" guidelines}

\subsection{Standard thin films with $L > 10$ nm}

A fundamental result of this paper is Eq. \eqref{extern}, which provides a possible  
microscopic estimate of the externally applied electric field needed to suppress superconductivity in thin films as a function of key physical parameters that depend on the material and/or the sample preparation. 
In addition to offering
a quantitative prediction of the critical field $E_{cr,ext}$ in agreement with experimental data, this formula presents new guidelines for materials design. 
For instance, the impact of structural disorder will influence both $\Delta$ and $\ell$. The influence of disorder on $\Delta$ parallels its effect on $T_c$, and is notably complex and dependent on the material and sample \cite{Ovadyahu1973,Setty2020}. In the standard s-wave Eliashberg theory, of course, the disorder does not affect $T_c$ and $\Delta$, but this is no longer true in the generalized Eliashberg theory. Furthermore, in the case of intense disorder, the spectral function and the Coulomb pseudopotential can change.
In some materials, the disorder will increase $T_c$ and, consequently, $\Delta$; in others, it will cause a decrease. In other materials, the effect can be non-monotonic, exhibiting either a dome or a minimum (for instance, for Pb-based alloys, $T_c$ exhibits a minimum as a function of disorder \cite{Setty2020}). Conversely, we anticipate that the mean free path $\ell$ will consistently decrease as disorder increases \cite{deGennes}.
Thus, Eq. \eqref{extern} predicts various scenarios where one should optimize the disorder dependence of the ratio $\Delta/	\ell$. For example, to reduce the critical field $E_{cr,ext}$, it is advisable to adjust the structural disorder to minimize the ratio $\Delta/\ell$.

We note that the value of $E_{cr,ext}$ also depends on the film thickness. Specifically, since $L\gg l$, the exponential in the formula becomes a small number compared to 1; therefore, the primary effect of decreasing $L$ is to lower the critical field $E_{cr,ext}$.

Finally, the effect of the EF penetration depth $l$ on the critical field is quite clear: once again, because $L \gg l$, the primary consequence of increasing $l$ will be a reduction in the critical field $E_{cr,ext}$. Generally, one would expect the screening penetration depth to be more significant for low-density superconductors. However, while this is evident from Eq. \eqref{extern}, it is certainly not straightforward to determine how to achieve this kind of fine-tuning in materials engineering, given the current lack of microscopic insight into the dielectric screening in the superconducting phase of metals.

\subsection{Ultra-thin films with $L < 10$ nm}
The above illustrative calculation for NbN thin films was done in a regime where the energy gap $\Delta$ is independent of the film thickness. However, several materials of common use, including Al and Pb, exhibit a strong dependence of their $T_c$ and $\Delta$, on the thickness $L$, when $L$ approaches the 2D limit \cite{zaccone,nostro,gold}. These effects may become important already at $L< 10$ nm, and the film thickness, which marks the onset of these effects, varies (increases) with the free-carriers concentration in the material. In most cases, the $T_c$ and hence $\Delta$ increase markedly as the film thickness $L$ decreases, as observed experimentally \cite{Scheffler,Arutyunov2019,AlTc}, although sometimes also a maximum in $T_c$ vs $L$ is observed \cite{lead2,doi:10.1126/sciadv.adf5500} with a regime where $T_c$ grows with $L$. The recently developed Eliashberg's theory corrected for quantum confinement \cite{nostro} can describe quantitatively these effects without free parameters \cite{nostro}.

The quantum confinement corrections to the Eliashberg theory amount to having a thickness-dependent Fermi energy, $E_F(L)$, and an electron DOS which is also thickness-dependent $N(\epsilon;L)$. The specific functional forms of these dependencies can be found in Refs. \cite{zaccone,nostro} and are summarized in Appendix A. All this implies that the energy gap $\Delta(L)$ entering Eq. \eqref{extern} for the critical EF magnitude will also become a function of the film thickness $L$. Therefore, for ultra-thin films, Eq. \eqref{cr} should be replaced by:
\begin{equation}
E_{cr,ext}=\frac{2 \Delta(L)/e\,\ell} {\frac{2l}{L}(1- e^{-L/(2l)})}, \label{extern_2}
\end{equation}
with an explicit dependence of $\Delta$ on $L$ due to quantum confinement effects.
Generally, knowing $\Delta(L)$ from theory \cite{nostro}, one can minimize the function $E_{cr,ext}(L)$ defined by the above Eq. \eqref{extern_2} to identify the thickness value corresponding to the lowest critical EF required to suppress superconductivity in this regime.

\section{Conclusions}
In conclusion, we have presented a possible route to a microscopic quantitative theory of electrostatic field-driven superconductivity suppression in thin films. The theoretical model treats the Cooper pair as an s-wave bound state (following the original description by Cooper \cite{Cooper}), which is subjected to a small but non-vanishing electric field within the film. Using the Schr\"{o}dinger equation solution for s-wave bound state dissociation through electric-field-assisted tunneling, we derive an expression for the critical or characteristic electric field magnitude inside the film to break the Cooper pairs, thereby suppressing superconductivity. This critical electric field is proportional to the ratio between the superconducting energy gap value $\Delta$ and the coherence length $\xi$, as shown in Eq. \eqref{crit}. The gap value is related to the $T_c$ value through the BCS relation $2\Delta/k_BT_c \approx 3.53$. However, in our framework, we computed it via the most accurate Eliashberg theory, which considers structural and disorder effects on the phonon spectral function for a specific material, achieving quantitative agreement with experimental data for the behavior of $T_c$ without any adjustable parameters.

Since the $T_c$ of the materials used in the experiments \cite{review} is comparable, so are the energy gaps $\Delta$. Regarding the coherence length $\xi$, it should be noted that the experiments are typically in the diffusive limit for many materials. Therefore, we need to consider the coherence length value obtained from Eq. \eqref{matt}, as the mean free path $\ell$ is always relatively small in real materials and of the same order of magnitude in all experimental cases studied in the literature \cite{review}. 
Generally speaking, $\xi$ is typically the same for most materials investigated experimentally to date \cite{review}; hence, the critical field value primarily depends on the energy gap value $\Delta$.

The outcome of the theory is given by Eq. \eqref{extern}, which provides a quantitative estimate of the critical field required to suppress superconductivity, as it relates to all key microscopic material- and sample-dependent parameters: energy gap $\Delta$, mean free path $\ell$, film thickness $L$, and the EF screening length in the superconducting phase, $l$.


Focusing on the 20-nm-thick NbN films, for which the theory provides a perfect quantitative prediction of $\Delta$, we used Eq. \eqref{extern} to compute the critical electric field required to break up the Cooper pairs and suppress superconductivity. Based on our Eliashberg theory calculation for realistic NbN films and considering the spatially decaying profile of the electric field within the sample, our theory predicts a critical field $E_{cr} \approx 1.45 \cdot 10^8$ V/m, which aligns well with the experimentally measured values of $\sim 10^8$ V/m \cite{review}. Combined with the underlying electrostatic model, these calculations provide a comprehensive understanding of the electric field-driven suppression of superconductivity in metallic thin films. This understanding accounts for the inhomogeneous distribution of the electric field within the sample due to dielectric screening in the superconducting phase. 


The theory is very general and extends beyond this specific calculation, as it applies to all thin films made of conventional materials. It also accounts for the dependence on film thickness due to quantum confinement in the ultra-thin film regime ($L < 10$ nm), where the energy gap varies significantly with thickness \cite{zaccone,nostro,gold}. 
This could explain experimental data related to measurements of films with varying thicknesses down to the quasi-2D limit. These new insights and understandings may enable us to deliberately adjust the supercurrent field effect in gating and nanoelectronic devices based on superconductors.

\appendix
\section{Electron confinement model and calculation details}
This Appendix summarizes how the electronic DOS and Fermi energy depend on the film thickness $L$ via the relations derived from the free-electron confinement model for thin films of Ref. \cite{zaccone} and quantitatively verified in comparison with experimental data in Ref. \cite{nostro}.
In the calculations presented in the main article, the film thickness effects over the DOS are practically negligible since they are set in only for ultra-thin films with thickness $L< 10$ nm. However, for completeness, the dependencies on $L$ are made explicit.

When the system is confined along one of the three spatial directions, such as in thin films, the DOS cannot be approximated by a constant but takes a different form \cite{zaccone,nostro}.
In this case, we have two different regimes depending on the film thickness $L$:
in the first confinement regime, when $L>L_{c}$ and $E_{F}>\varepsilon^{*}$, the DOS has the following form

$N(\varepsilon)=N(0)C
[\vartheta(\varepsilon^{*}-\varepsilon)\sqrt{\frac{E_{F}}{\varepsilon^{*}}}\frac{|\varepsilon|}{E_{F}}+
\vartheta(\varepsilon-\varepsilon^{*})\sqrt{\frac{|\varepsilon|}{E_{F}}}]$

\noindent where $C=(1+\frac{1}{3}\frac{L_{c}^{3}}{L^{3}})^{1/3}$, $\varepsilon^{*}=\frac{2\pi^{2}\hbar^{2}}{mL^{2}}$, $L_{c}=(\frac{2\pi}{n_{0}})^{1/3}$,
$m$ is the electron mass, $L$ is the film thickness, $n_{0}$ is the density of carriers and $E_{F,bulk}$ is the Fermi energy of the bulk material. In this case, it is possible to demonstrate \cite{zaccone} the following relations:
\begin{align}
E_{F}&=C^{2}E_{F,bulk}\\
N(E_{F})&=CN(E_{F,bulk})=CN(0)\\
N(E_{F,bulk})&=\frac{V(2m)^{3/2}}{2\pi^{2}\hbar^{3}}\sqrt{E_{F,bulk}}.
\end{align}
In the second confinement regime, $L < L_c$, the DOS has a new, linear dependence on the energy, in contrast with the standard square-root dependence \cite{zaccone}.

To summarize, in this version of the Eliashberg theory, four things have been modified to account for the effect of free-electron confinement in thin films:

a) the DOS will no longer be a constant but a function of energy.
We removed the factors $C$ because we put this factor in the renormalization of the electron-phonon interaction, so the density of states
that we put in the Eliashberg equations in the first confinement regime $L>L_c$ is
\begin{equation}
N(\varepsilon)=[\vartheta(\varepsilon^{*}-\varepsilon)\sqrt{\frac{E_{F}}{\varepsilon^{*}}}\frac{|\varepsilon|}{E_{F}}+
\vartheta(\varepsilon-\varepsilon^{*})\sqrt{\frac{|\varepsilon|}{E_{F}}}].
\end{equation}

b) the electron-phonon interaction will be renormalized to have a new $\lambda(L)=C\lambda^{bulk}$ in a way to scale the electron-phonon spectral function without changing its shape. We moved the factor of the normal DOS $C$ inside the definition of electron-phonon coupling as in the Coulomb pseudopotential. Of course, the reason for this choice is only pedagogic because, in this way, we can justify the use of the Allen-Dynes equation \cite{Dynes} for $T_{c}$, which is a crude but effective approximation of the numerical solution of the Eliashberg equations.

c) the value of the Fermi energy will be renormalized in the following way: $E_{F}(L)=C^{2}E_{F,bulk}$.
Furthermore, in the Eliashberg equations, it is $W=E_{F}/2$ in the symmetric case discussed above.

d) the Coulomb pseudopotential changes (also the Fermi energy in the definition, changes):

$\mu^*(\omega_c,L)=\frac{C\mu_{bulk}}{1+\mu_{bulk}\ln(E_{F}/\omega_{c})}$
\noindent where $\mu_{bulk}=\frac{\mu^{*}_{bulk}(\omega_c)}{1-\mu^{*}_{bulk}(\omega_c)\ln(E_{F,bulk}/\omega_{c})}$.

In the second confinement regime, when $L<L_{c}$ and $E_{F}<\varepsilon^{*}$ \cite{zaccone}:
\begin{equation}
    N(\varepsilon)=C'N(0)\sqrt{\frac{E_{F}}{\varepsilon^{*}}}\frac{\varepsilon}{E_{F}}
\end{equation}

\noindent where

$N(\varepsilon=E_{F})=C'N(0)$

$E_{F}(L)=\frac{\hbar^{2}}{m}\sqrt{\frac{(2\pi)^{3}n_0}{L}}=C'^{2}E_{F,bulk}$

$C'=\frac{2}{6^{1/3}}\sqrt{\frac{L}L_c}$.\\

In this confinement regime ($L < L_c$), the DOS is given by \cite{zaccone}:

$N(\varepsilon)=\sqrt{\frac{E_{F}}{\varepsilon^{*}}}\frac{|\varepsilon|}{E_{F}}$,

\noindent and the factor $C'$ goes to renormalize the
electron-phonon coupling and the Coulomb pseudopotential as follows:
\begin{equation}
\lambda(L)=C'\lambda^{bulk} ,~~~ \mu^{*}(\omega_c,L)=\frac{C'\mu_{bulk}}{1+\mu_{bulk}\ln(E_{F}/\omega_{c})}.
\end{equation}


By recalling that in the Eliashberg equations, the reference energy is the Fermi energy taken as the zero of the energy, in the program that numerically solves the Eliashberg equations, the DOS has been rescaled in the following way
(by also putting care that the DOS is continuous for $\varepsilon=\varepsilon^{*}$). When $L>L_{c}$ and $\varepsilon^{*}<E_{F}$:
\begin{equation}
N(\varepsilon)=[\vartheta(\varepsilon^{*}-\varepsilon)\sqrt{\frac{E_{F}}{E_{F}-\varepsilon^{*}}}(1-\frac{|\varepsilon|}{E_{F}})+
\vartheta(\varepsilon-\varepsilon^{*})(1-\sqrt{\frac{|\varepsilon|}{E_{F}}})].
\end{equation}

Instead, when $L<L_{c}$ and $\varepsilon^{*}>E_{F}$:
\begin{equation}
N(\varepsilon)=\sqrt{\frac{E_{F}}{\varepsilon^{*}}}(1-\frac{|\varepsilon|}{E_{F}}).
\end{equation}
Finally in our case for the illustrative calculation on the example of NbN, we have used $\lambda_{bulk}=1.46$, $\mu_{bulk}^*(\omega_c)=0.2522$ for $\omega_c=180$ meV, $E_{Fbulk}=12.418$ eV, $n_0=0.197\cdot 10^{30}$ $m^{-3}$, $l=0.4$ nm and $v_F=2.1\cdot 10^6$  m/s. \\

\subsection*{Acknowledgments} 
A.Z. gratefully acknowledges funding from the European Union through Horizon Europe ERC Grant number 101043968 ``Multimech'', from the US Army Research Office through contract nr. W911NF-22-2-0256, and from the Nieders{\"a}chsische Akademie der Wissenschaften zu G{\"o}ttingen in the frame of the Gauss Professorship program.
F.G. acknowledges the EU’s Horizon 2020 Research and Innovation Framework Programme under Grants No.
964398 (SUPERGATE), No. 101057977 (SPECTRUM), and the PNRR MUR project PE0000023-NQSTI for
partial financial support. A.B acknowledges MUR-PRIN 2022 - Grant No. 2022B9P8LN
- (PE3)-Project NEThEQS “Non-equilibrium coherent
thermal effects in quantum systems” in PNRR Mission 4 -
Component 2 - Investment 1.1 “Fondo per il Programma
Nazionale di Ricerca e Progetti di Rilevante Interesse
Nazionale (PRIN)” funded by the European Union - Next
Generation EU and the PNRR MUR project
PE0000023-NQSTI and CNR project QTHERMONANO.

\bibliographystyle{apsrev4-2}

\bibliography{refs}

\end{document}